\documentclass[11pt, a4paper]{paper}
\usepackage{geometry}                
\usepackage{graphicx}
\usepackage{amssymb}
\usepackage{epstopdf}
\usepackage{authblk}
\usepackage[applemac]{inputenc}
\title{A model for discussing entropy and time reversibility}
\author{Tommaso Castellani\thanks{tommaso.castellani@cnr.it}}
\affil{Science Communication and Educaton, \\ the National Research Council of Italy, Rome}

\begin{document}
\maketitle
\begin{abstract}
In this article we discuss a model used to introduce the concept of entropy with secondary school students. It can be used to discuss with students the reversibility of time, the tendency towards homogeneity and the link between probability theory and second law of thermodynamics. The model is useful to introduce crucial epistemological issues and helps student to understand the deep connection between the macroscopic and the microscopic.
\end{abstract}
Entropy is a fascinating concept, as well as one of the hardest to grasp for a secondary school student.
Whereas the claim that ``the entropy of an isolated system never decreases'' is an elegant and apparently student-friendly formulation of the Second Law of Thermodynamics, the deep understanding of what this statement actually means is not at all straightforward.

In this paper we propose a model to discuss with students the concept of entropy by the point of view of statistical physics. This approach not only clarifies the connection from the macroscopic to the microscopic, one of the core issues of thermodynamics, but gives also occasion to discuss more general and fascinating questions as the reversibility of time.

For secondary school students, thermodynamics is challenging also because it entails a change of paradigm with respect to the newtonian physics. Whereas the latter is characterized by prescriptive laws as ``a system in this condition will do this'', thermodynamics laws have the form of ``prohibitions'', as ``a system in this condition will \textit{not} do this''. The model presented in this paper helps students to understand the origin and the meaning of this crucial epistemological issue.

Before introducing the model, we present some preliminary experiences that can be simply set up and discussed in the classroom, in order to introduce the topic and discuss the emerging questions. The teacher can start from the questions of the students, following an inquiry-based approach, and encouraging the production of students' own hypothesis and explanations.

\part*{Preliminary experiences}
\subsubsection*{Time reversibility}

The following experiment is usually introduced to demonstrate the microscopic motion of the matter and the relationship between the temperature and the average speed of the molecules. We propose a further discussion starting from the same experience.

In two glasses of water at different temperatures we throw a drop of coloured ink. The experience shows that the drop undergoes a process of diffusion, until the water is uniformly coloured. In the hottest glass the diffusions happens more rapidly, and this demonstrates that the temperature corresponds to molecules moving faster in average. 

For our purpose, we can use this experience asking students another question: is it possible that, at a certain point, the motion of the molecules makes the original ink drop to recreate? I.e. is the process \textit{reversible}?

Generally students agree it is not, since ``it never happens that a homogeneously coloured liquid is transformed into an isolated coloured drop''. We can ask students to make examples of other irreversible phenomena, and to think to instances of reversible ones. Since the characterization of some phenomena can lead to disputes, we can discuss how to define reversibility. At this stage it is not important to be rigorous, since we aim to stimulate students to elaborate their own ideas. 

We gradually lead the students to ask themselves what are the possible reasons for time irreversibility.

\subsubsection*{Homogeneity}

Together with the problem of reversibility, it came out the notion of ``homogeneity''. From the ink drop experiment it seems that homogeneity is a sort of ``final state'' which is not going to evolve any more.

In order to discuss this problem, we can introduce a further experiment. Two balls of plasticine of different colours are given to two students, who are asked to exchange some small quantity of plasticine with each other. The exchange is made about every minute, but the time is not fixed (i.e. it is not measured in any way), nor is the quantity of plasticine to be exchanged, which can vary time by time. This is actually a random process. The question is: what happens after a long time? Usually students agree that the colours of each ball will be both intermediate colours between the two initial ones, but they are divided on two hypothesis:
\begin{enumerate}
\item if we wait enough time, the two balls will take a perfectly identical colour;
\item also after a long time, the two colours will never be perfectly identical.
\end{enumerate}
The supporters of hypothesis 2 usually argue that, being the time of mixing and the quantity of matter exchanged not fixed, there is not any reason for the colours not to be different. The discussion can be carried forward while the two students continue to exchange plasticine from their balls. After about half an hour, everybody can observe that the colours are perfectly identical. This is also rather shocking for the supporters of hypothesis 2.

The question is: why randomness produced homogeneity?

\part*{The model}
The two previous experiments raised two questions that are immediately understandable and engaging for the students, while at the same time very difficult to solve. Both questions are central in the introduction of entropy, and stimulate students to think to the main issues of second law of thermodynamics. 

In order to introduce our model, we make an example that encompasses the two previous experiences. Imagine to have a glass half-filled with wine in the lower half and half-filled with water in the upper part, with a thin film that separates the two parts. What happens when the film is removed? Is this process reversible? 

This experiment---which is actually a mental experiment, since it is very difficult to realize in practice the a removable film which divides the glass perfectly in two halves---can be represented with a model. It is very important to discuss the meaning of the word ``model'', so crucial in science and so little addressed in school curricula. We must focus the attention on the fact that the model, which can be a material model (e.g. a molecule made by plastic balls) or an abstract model (e.g. a mathematical model), is a construction which aims to isolate and study \textit{some specific aspects} of a phenomenon.

Our model aims to reproduce the probabilistic aspects of the mental experiment of the glass. We want to represent a sample of $N=40$ molecules of the liquids of our glass, 20 of liquid A and 20 of liquid B (we can continue to speak of water and wine if we prefer, provided we explain students that we will call ``molecules of wine'' the ``particles'' of wine, whatever they are---this may seem not at all rigorous, but is not after all so different from the way physicists speak of ``spins'' in magnetic materials). We should specify that the two liquids have the same density, so that there are no physical reasons for them to separate.

\begin{figure}[htbp] 
\begin{center} 
\includegraphics[width=12cm]{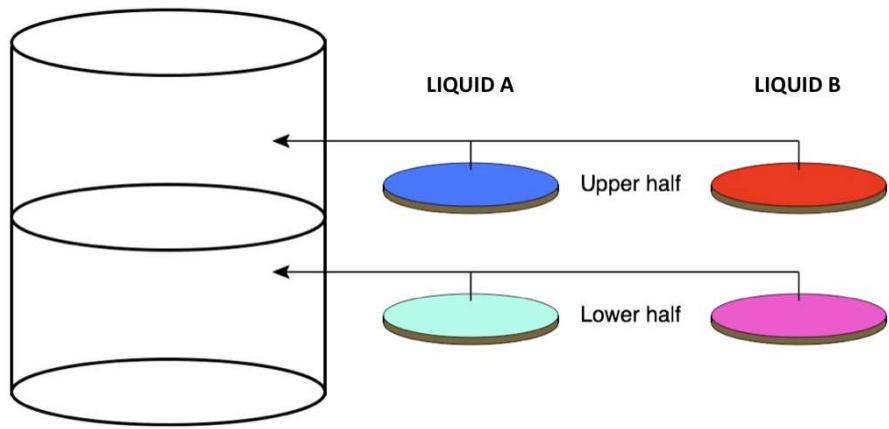} 
\end{center} 
\caption{The model: each cardboard disk represents a molecule. The colours represent both the different liquid and the different half of the glass in which the molecule is.} 
\label{fig1}
\end{figure}

We represent the 40 molecules with cardboard disks of different colours.  The model aims to focus on the position of the molecules inside the glass, and in particular in what half of the glass the molecule is, whether in the upper half or in the lower half. In order to represent this feature, we assign to the two sides of the cardboard disks the two halves of the glass, distinguishing them by marking with different colours (see Fig.\ref{fig1}). We will therefore have:
\begin{itemize}
\item Liquid A molecules:
\begin{itemize}
\item Side 1 (blue): liquid A molecules in the upper half of the glass
\item Side 2 (cyan): liquid A molecules in the lower half of the glass
\end{itemize}
\item Liquid B molecules:
\begin{itemize}
\item Side 1 (red): liquid B molecules in the upper half of the glass
\item Side 2 (magenta): liquid B molecules in the lower half of the glass
\end{itemize}
\end{itemize}

The molecules are put on the bottom of a box, that must be large enough to contain all the 40 cardboard disks on its bottom surface. Each disk will show a side. Depending on the colour, we know what molecule and in what half of the glass each disk represents.

Let's see now how the model is used to simulate the time evolution of the system. 
We want to fix the initial state with all liquid A molecules in the upper half of the glass and all liquid B molecules in the lower half. That corresponds, in our model, to putting all the A molecules on the blue side and all the B molecules on the magenta side. We open the box and, on the bottom of it, we put all the 40 molecules according to this condition.
This represents the initial state of our previous mental experiment: a glass in which water and wine are well separated. 

At this point we close the box and we shuffle it for a certain amount of time. This represent the thermal motion inside the glass after the removal of the film. We ask students how do they expect to find the molecules after, say, ten seconds of shuffling. Almost everybody say that the molecules will be ``mixed''. We open the box, the students count the molecules and see that for each liquid they are divided almost in half between the two colours, i.e. the two halves of the glass. 
\subsubsection*{Question 1: the reversibility of time}
We ask students: if we shuffle the box some other times, will the model come back to the initial condition---with separated liquids---sooner or later? Some students answer ``yes”, some are undecided, some others say that it is ``actually impossible”. 

The discussion on the word ``impossible” generates a cognitive conflict---how an event can be impossible if there is no physical law preventing it? We will use this conflict to introduce the topic of statistical improbability. 
With some basic mathematics it is easy to calculate the probability of utter separation of the two liquids of our model, which is:
$$
P_{separation}={\left(\frac12\right)}^{40}
$$
How much is it? We can estimate with the students that if each shuffling takes 10 seconds and we shuffle all day and night, we will see the separation in average one time every 350,000 years. This is quite shocking!

This discussion answers the question of the origin of irreversibility in the examples of ink drop and plasticine: the irreversibility is due to the extreme improbability of the ``reverse'' phenomenon. Although there is no physical reason for the 40 molecules of our model to be again in the initial setting (20 of liquid A on blue side-upper half and 20 of liquid B on magenta side-lower half), this condition is extremely unlikely, so unlikely that we will actually never see it.

We must remark to students that our example involved a sample of 40 molecules. What if we extend the model to a larger number of the molecules, which in the entire glass may reach the order of magnitude of Avogadro constant? With $10^{23}$ molecules, the probability of getting back to the initial state is so small that we can call this event ``impossible'', though not in the traditional (logical or mathematical) meaning.

At the end of the discussion, it is useful to stress once more to students that the proposed model is not a realistic description of the real glass, but it deals with the same probabilistic problems, shown very easily and in an immediate way. It shows as well the origin of the epistemological difference of thermodynamic laws, which appear as ``prohibitions'' rather than ``prescriptions'', as for instance is the dynamics laws. The students should see how the evolution of the system at microscopical level is driven by probabilistic laws, and this changes the nature of the statements at the macroscopic level.
\subsubsection*{Question 2: macroscopic and microscopic states}
The distinction between ``macroscopic state” and ``microscopic state'' is another very important concept that can be introduced by means of our model. We can define macroscopic state as the appearance of the glass/box at the human scale: e.g. ``a glass with a homogeneous mixture of water and wine'', or in our model ``both type of disks are on both sides with no prevalent colours''. The microscopic state is the description of the individual positions of each molecule: in our model, that means to label the molecules and say ``No. 1 is in the blue side, no. 2 is in the red, etc''.

\begin{figure}[htbp] 
\begin{center} 
\includegraphics[width=12cm]{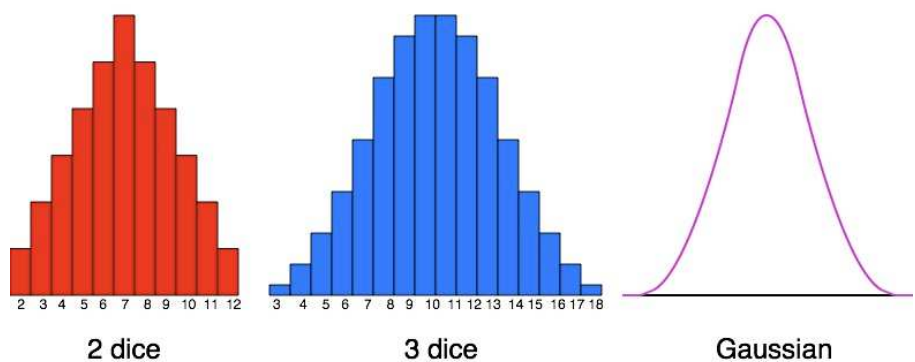} 
\end{center} 
\caption{The correspondence between ``macroscopic states'' (sum of dice values, on $x$ axis) and number of ``microscopic states'' (number of combinations producing that given sum, on $y$ axis). In the picture, different scales of $y$ axis are used in order to keep students' attention on the transformation of the shape and not on the numeric values.} 
\label{fig2}
\end{figure}

An interesting way to introduce macroscopic and microscopic states is to use dice. We take two dice, and we call ``macroscopic state” the sum of the values of the two dice, while we call ``microscopic state” the couple of numbers describing each dice. For each sum, we count the number of combinations of the two dice which produce that sum. The sum 2, for example, is only obtained when both dices show a 1. The sum 7 is the one with the largest number of possible combinations, since it can be obtained in 6 ways (1+6, 2+5, 3+4, 4+3, 5+2, 6+1). 
The macroscopic states (i.e. sums) with the largest number of corresponding microscopic states (i.e. single dice values) are the most likely. We can draw a histogram of the number of microscopic states corresponding to each macroscopic state (see Fig.\ref{fig2}): students can clearly see that there are macroscopic states more likely than others. We can draw the same histogram increasing the number of dice: we note that at the ``tails” of the histogram remains small, while the central part increases a lot (this is the way towards the Gaussian curve) (Fig.\ref{fig2}). 

It is straightforward for students to understand that, for a number of dice that approximates to $10^{23}$, we will have very very unlike macroscopic states and a tiny central zone with macroscopic states which are corresponding with a very large number of microscopic states.

Coming back to our model, we can make analogue considerations. The macroscopic state ``all liquid A molecules in the upper half, all liquid B molecules in the lower half'' corresponds to only 1 single microscopic state, in which all the A molecules are on blue side and all the B molecules on magenta side. As we did for dices, we can order the macroscopic states and proceed to consider the macroscopic state ``all except one liquid A molecules in the upper half, all liquid B molecules in the lower half''. The microscopic state corresponding to this macroscopic state can be obtained by turning on the cyan side one of the liquid A molecules. Nevertheless, having 20 water molecules, we can indifferently chose one of the 20: so we have 20 possibilities for making this state. The macroscopic state ``all except one liquid A molecules in the upper half, all liquid B molecules in the lower half'' has 20 corresponding microscopic states. 
Going on, the macroscopic state all except one liquid A molecules in the upper half, all liquid B molecules in the lower half'' has 400 ($20\cdot20$) corresponding microscopic states, and so on. Each of this macroscopic states has its symmetric if we invert A and B.

Students can see that the more we approach the ``homogeneous'' state, the more microscopic states correspond to a single macroscopic state, and that their number increases exponentially. It must be said what we call ``homogeneous'' state is actually a number of different macroscopic states, i.e. all the states in which the number of molecules in the two sides is \textit{almost} the same (this is what students actually see when they count the molecules of the mixed state of the model, which do not need to be divided \textit{exactly} in half).
As in the case of dice, we have ``tails” of very unlikely events, and a central part of a histogram which contains the majority of microscopic states. This central part if what we actually call ``homogeneous state”.

This discussion explains why the homogeneous condition is the natural outcome of the evolution of the process: it is much more likely than all the other conditions. 
\subsubsection*{Question 3: defining entropy}
We can finally define \textit{entropy} of a macroscopic state the number of microscopic states associated to it. Depending on the mathematical background of the students, we can or can not define it rigorously as the logarithm of this number. We may as well multiply by the Boltzmann constant $K$ and give the precise definition.

In any case, the experience with the box model has given to the students an immediate and easy to understand overview of the physical and mathematical questions behind the notion of entropy. The formulation of the second law of thermodynamics as a non-increasing of entropy in an isolated system is now more easily understandable as a consequence of probability of macroscopic states.

Sometimes, books use to say that entropy is a ``measure of disorder''. It may be useful to clarify what disorder means: is the homogeneity considered disorder by students? We can ask them to say if it is more ordered a ball of plasticine of a perfectly homogeneous colour or another one with some random stripes of two colours still not well mixed. Considering the common language, many students will probably answer that the first one is more ordered, while physicists considered homogeneity more ``disordered''. We suggest not to use this term as it can be misleading, but if used, it must be stressed that its sense may be different from the sense it has in the common language, as---by the way---for most physical terms that have a common meaning.

\part*{Conclusions}
We presented a model to discuss the notion of entropy from a statistical physics point of view with secondary school students. 
The model offers the opportunity of discussing the epistemological features of thermodynamics. Epistemological aspects has long been considered crucial in science education \cite{cini1995, osborne2003}: this model allows to deal with them in a comprehensible and stimulating way. The model can be used together with the experiences proposed in this paper, as well as with other models for related phenomena, like Atkins models on heat diffusion \cite{atkins1984}, or other more complex educational pathways related to the second law of thermodynamics \cite{kincanon2013, castellani2013a}. We also suggest to start from this approach to discuss the many common students' misconceptions on microscopic motion of the matter and kinetic theory \cite{pathare2010}, often present also in textbooks \cite{tarsitani1996}.

Our model has been proved to be successful in many different contexts, from the curricular lesson to specific laboratories on modern physics themes. We always experienced a great involvement of the students and a stimulating debate.

\end{document}